\documentclass{blois}

\newcommand{\Photo}{}
\usepackage{wrapfig}

\begin{document}
\vspace*{4cm}
\title{The science of ultra-high energy cosmic rays after more than 15 years of operation of the Pierre Auger Observatory}

\author{Olivier Deligny\textsuperscript{*},
for the Pierre Auger Collaboration\textsuperscript{**}}

\address{{\bf \textsuperscript{*}} Laboratoire de Physique des 2 Infinis Ir\`ene Joliot-Curie (IJCLab)\\
CNRS/IN2P3, Universit\'{e} Paris-Saclay, Orsay, France
\\
{\bf \textsuperscript{**}} Full author list: \href{https://www.auger.org/archive/authors\_2022\_05.html}{https://www.auger.org/archive/authors\_2022\_05.html}
}

\maketitle\abstracts{Ultra-high energy cosmic rays (UHECRs) have been studied with the data of the Pierre Auger Observatory for more than fifteen years. An essential feature of the Observatory is its hybrid design: UHECRs are detected through the observation of the associated extensive air showers (EASs) with different and complementary techniques. The analyses of the multi-detector data have enabled high-statistics and high-precision studies of the UHECR energy spectrum, mass composition and distribution of arrival directions. The resulting science of UHECRs is summarized in this contribution. While no discrete source of UHECRs has been identified so far, the extragalactic origin of the particles has been recently confirmed from the arrival directions above 8~EeV, and the ring is closing around nearby astrophysical sites at higher energies. Also, the established upper limits on fluxes of UHE neutrinos and photons have implications on multi-messenger studies and beyond-the-Standard-Model (BSM) physics.}

\section{The Pierre Auger Observatory: An astroparticle laboratory}

The Pierre Auger Observatory~\cite{PierreAuger:2015eyc} has been on the astroparticle scene for over fifteen years now. It's a mature player in this field, yet it retains all its vitality to provide data whose richness sheds light not only on the origin of UHECRs, but also on high-energy hadronic interactions, multi-messenger astrophysics, BSM physics and atmospheric electricity phenomena. Only the surface of some of these topics is going to be scratched in this brief overview. 

Originally conceived as a detector covering an area of 3000~km$^2$ dedicated to the study of cosmic rays with energies in excess of 10~EeV thanks to an array of particle detectors surrounded by fluorescence telescopes, numerous enhancements, ranging from the design of more sensitive triggers to the construction of additional detectors, have enabled the Observatory to extend its range of applications. Nested arrays of particle detectors allow us nowadays to measure the energy spectrum of cosmic rays and large-scale anisotropies in their arrival directions above a few dozens of PeV; radio-detection of EASs as well as underground muon detectors provide new insights on the development of the cascades; uses of hybrid data streams allow for lowering the threshold to search for high-energy photons; and the good old canonical particle detectors and fluorescence telescopes turn out to be sensitive to down-going transient luminous events and to ELVES during special classes of thunderstorms. A fair amount of data that fed several generations of PhD students. 

After all these years of loyal service, the Observatory needed a major upgrade. Currently being commissioned, it capitalizes over one of the main breakthroughs concerning the nature of UHECRs: their bulk is composed of nuclei that get heavier with energy above $1.5~$EeV.\cite{PierreAuger:2014sui} This measurement sealed the fate of a paradigm, long considered as \textit{the} benchmark model, based on an exclusive composition of protons. The motivation for this paradigm was twofold: on the one hand, it provided a unique explanation for the presence of an ankle and of a suppression in the energy spectrum of UHECRs, and on the other hand, it relied on the minimal assumption that extreme acceleration mechanisms operated on the particles present in greatest numbers in interstellar and intergalactic media since the epoch of reionization, namely hydrogen nuclei. Despite the attractiveness of this paradigm, ultra-high-energy astrophysics as revealed by the Observatory is consistent with a different mass composition, also including heavier nuclei. While the presence of these nuclei is no longer open to debate given the numerous observables consistent with composition getting heavier with energy, the exact composition of the elements remains uncertain beyond $\simeq 30~$EeV. The upgrade of the Observatory has therefore been designed to characterize the exact composition of these ultra-high-energy nuclei as precisely as possible, using particle detectors capable of measuring the muon and electromagnetic components of the EAS in different ways. Analyses over the next decade will thus aim to determine the balance between medium ($\sim$CNO), heavy ($\sim$Si) and very heavy ($\sim$Fe) nuclei at ultra-high energies, which will help identify the nature of the shocks responsible for particle acceleration in given astrophysical environments.

One of the other main breakthroughs from the Observatory is the energy estimation of EAS in a calorimetric way even for events sampled only by ground-level particle detectors, which, as far as data prior to the upgrade are concerned, consist of water-Cherenkov detectors operating with a quasi-permanent duty cycle arranged on a triangular grid of 1500~m spacing over 3,000~km$^2$, except for denser infilled areas for lowering the energy threshold. The grid is overlooked by 27 fluorescence telescopes that detect the faint UV light emitted by nitrogen molecules previously excited by the charged particles from the EAS. This technique allows for performing EAS calorimetry by mapping the ionization content along the tracks of the cascades, and for measuring the primary energy on a nearly model-independent basis. The telescopes can only operate during dark nights with low moonlight with a field of view free of clouds ($\sim 15\%$ of the time). Online and long-term performances of the detectors and data quality are monitored continuously. The assessment of the energies of the events observed with the particle detectors makes use of a subset of events detected simultaneously by the fluorescence telescopes.\cite{PierreAuger:2008rol} This ``hybrid'' approach allows therefore a calorimetric estimate of the energy for events recorded during periods when the telescopes cannot be operated. Otherwise, assumptions about the primary mass and the hadronic processes that control the cascade development would be mandatory, which proves to be a difficult task as the primary mass on an event-by-event basis is unknown and the centre-of-mass energy reached at the LHC corresponds only to that of a proton of $\simeq 100~$PeV colliding with a nitrogen nucleus.

\section{Energetics, mass composition and arrival directions of UHECRs}

\begin{wrapfigure}{L}{8 cm}
{\includegraphics[width=0.45\textwidth]{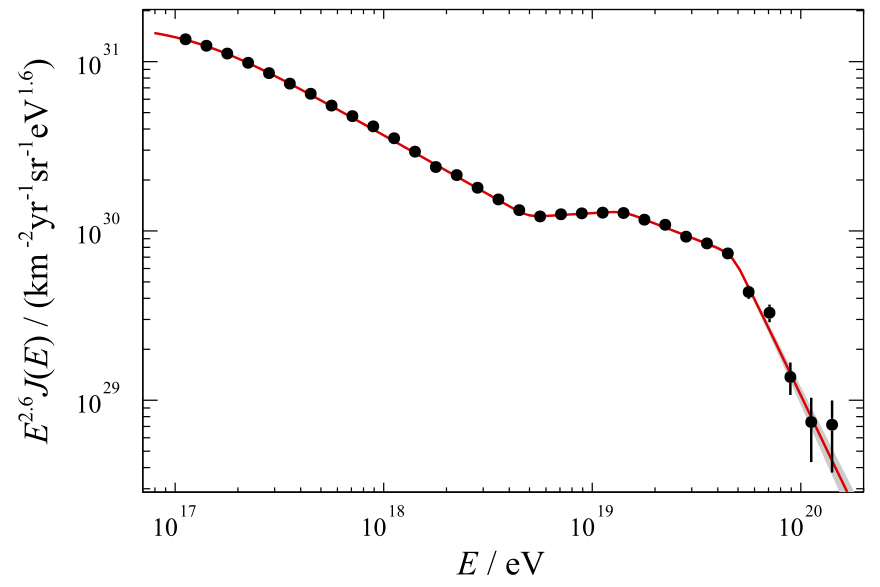}}
\caption{Energy spectrum of UHECRs, scaled by $E^{2.6}$.}
\label{fig:spectrum}
\end{wrapfigure}

The measurement of the all-particle energy spectrum above $100~$PeV is emblematic of the power of using multiple detectors: it has been performed by using the data from the two arrays of water-Cherenkov detectors complementary for covering the energy range down to $100~$PeV up to the highest energies, and from the fluorescence detectors for the energy calibration.\cite{PierreAuger:2021hun} The two spectra, in agreement within uncertainties, are combined into a single one shown in  Fig.~\ref{fig:spectrum} (scaled by $E^{2.6}$), taking into account the systematics of the individual measurements. Besides the ankle (the hardening at $\simeq 5~$EeV) and suppression (steepening at $\simeq 50~$EeV), spectral features established in the early years of the Observatory, the second-knee feature is observed to extend over a wide energy range (not fully covered in this measurement~\footnote{A measurement subsequent to the overview given at this conference, making use of an even denser infilled area of water-Cherenkov detectors, covers a wider energy range and allows for full observation of the second knee.}) while a steepening at $\simeq 10~$EeV, dubbed as the instep feature,\cite{PierreAuger:2020qqz} has been uncovered thanks to the exposure and energy resolution reached at the Observatory. It is to be noted, and this is one important legacy of the Observatory, that \textit{the all-particle energy spectrum shown in Fig.~\ref{fig:spectrum} does not depend on any assumption on the primary mass of UHECRs.}

\begin{figure}[h]
\centering
\includegraphics[width=0.75\textwidth]{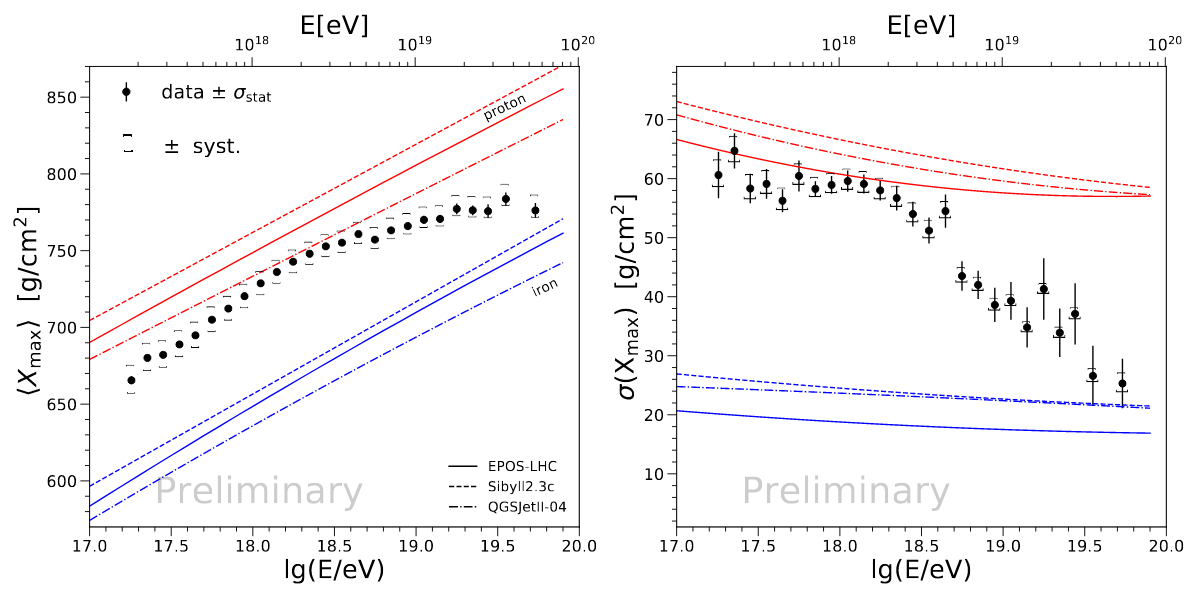}
\caption{Measurements of $\langle X_{\mathrm{max}}\rangle$(left) and $\sigma(X_{\mathrm{max}})$ (right) compared to the predictions for proton and iron nuclei of the hadronic models EPOS-LHC, Sibyll 2.3c and QGSJetII-04.}
\label{fig:xmax}
\end{figure}

The measurement of the depth of the shower maximum, $X_{\mathrm{max}}$, is the most robust mass-sensitive EAS observable. The mean and fluctuations of $X_{\mathrm{max}}$ data are shown in Fig.~\ref{fig:xmax}, as obtained from direct measurements with the fluorescence detectors.\cite{PierreAuger:2019phh} Between $\simeq 150~$PeV and $2~$EeV, $\langle X_{\mathrm{max}}\rangle$ increases by around 77~g~cm$^{-2}$ per decade of energy. This is larger than expected for a constant mass composition ($\simeq 60~$g~cm$^{-2}$ per decade) and thus indicates that the mean primary mass is becoming lighter over this energy range. Finally, above $2~$EeV, the rate of change of $\langle X_{\mathrm{max}}\rangle$ becomes significantly smaller ($\simeq 26$~g~cm$^{-2}$ per decade). This signals that the composition gets heavier. The fluctuations of $X_{\mathrm{max}}$ start to decrease above the same energy, $2~$EeV, being rather constant below. Altogether, the mean and fluctuations of $X_{\mathrm{max}}$ data are consistent with mass groups getting heavier  beyond the ankle energy and taking over one after the other so that the all-particle flux gets dominated by one specific mass group as the energy increases.  This behavior is confirmed by the composition fractions obtained from fitting templates of four mass groups to the $X_{\mathrm{max}}$  distributions.\cite{PierreAuger:2014gko}

\begin{figure}[h]
\centering
\includegraphics[width=0.49\textwidth]{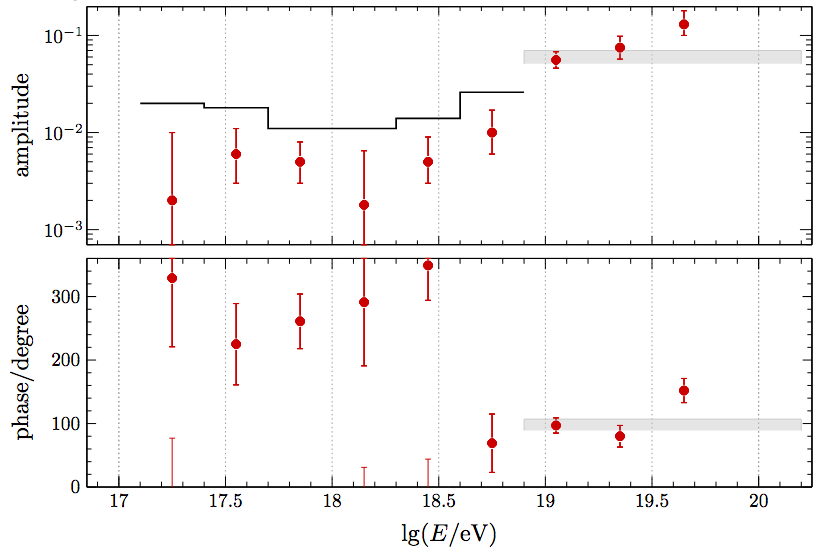}
\includegraphics[width=0.49\textwidth]{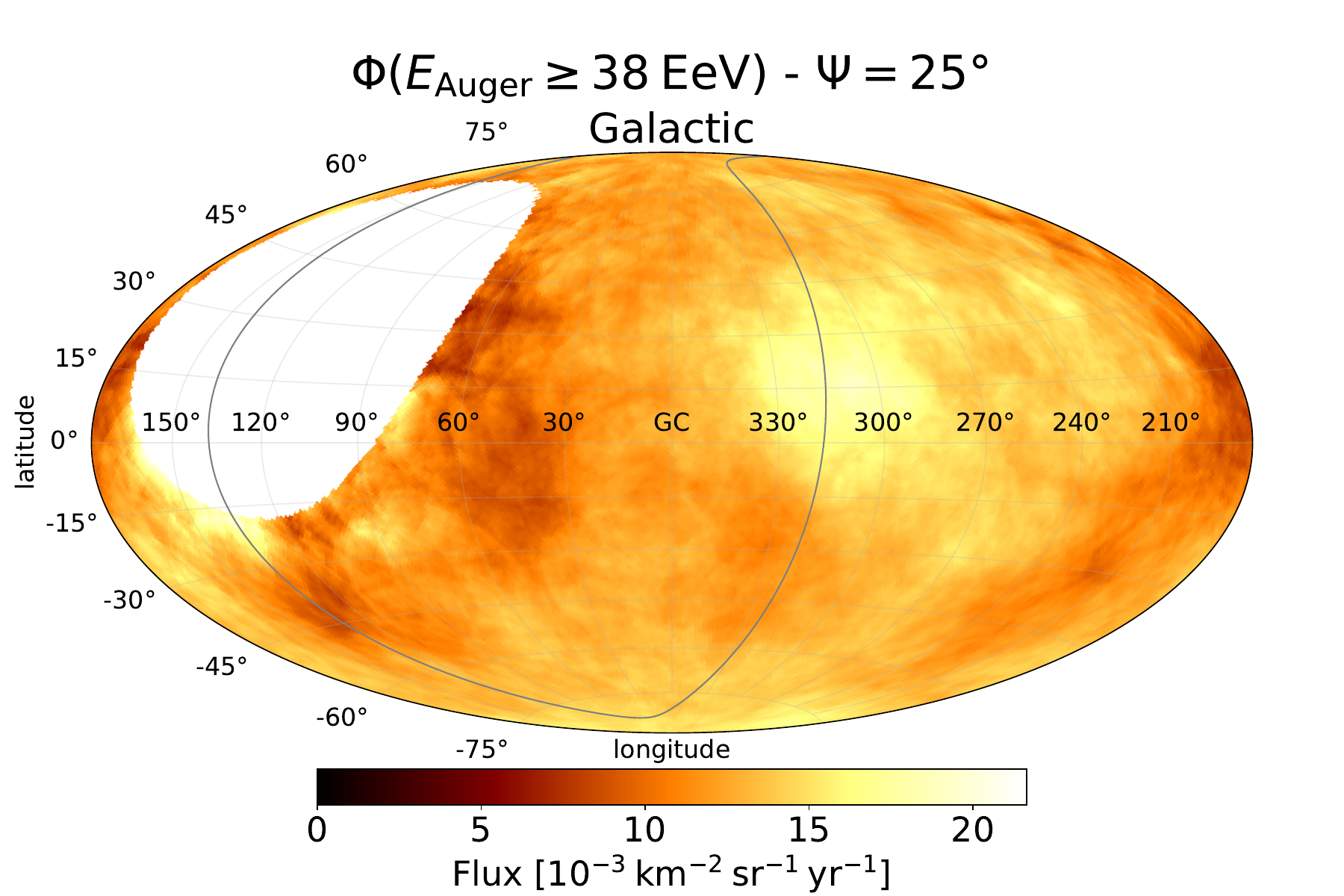}
\caption{Left: Amplitudes and phases of first harmonic in right ascension as a function of energy. Data from. Right: Map in Galactic coordinates of the total intensity above 38~EeV filtered at an angular scale of $25^\circ$. The supergalactic plane is indicated as the dashed line.}
\label{fig:largescale}
\end{figure}

Magnetic deflections spread the source fluxes over large solid angles. Anisotropy contrasts are thus expected to be small. The patterns are usually expressed in the reciprocal space. As the water-Cherenkov detectors operate almost uniformly with respect to sidereal time thanks to the rotation of the Earth, the most commonly used technique is the analysis in right ascension ($\alpha$) only, through harmonic analysis of the counting rate within the declination ($\delta$) band defined by the field of view of the Observatory. The corresponding amplitudes and phases are shown in Fig.~\ref{fig:largescale} (left panel).\cite{PierreAuger:2020fbi} Upper limits at 99\% C.L. at the percent level are shown as the black lines below $8~$EeV, where none of the measured amplitudes are significant. However, a consistency of phases is observed with a change at $\simeq 2~$EeV, consistency potentially indicative of a real underlying anisotropy.\cite{Edge:1978rr,Deligny:2018blo} It is worth noting that the phases below $\simeq 2$~EeV line up with the right ascension of the Galactic Center ($\simeq 268^\circ$). Above $8~$EeV, an anisotropy signal with a direction of maximum intensity pointing towards $(\alpha,\delta)=(\simeq 99^\circ,-23^\circ)$ in Galactic coordinates has been discovered;~\cite{PierreAuger:2017pzq} it is depicted as the gray band in Fig.~\ref{fig:largescale}. This signal can be fairly reproduced for sources that mimic the large-scale distribution of the baryonic matter. For the infrared-detected galaxies in the 2MASS Redshift Survey (2MRS) catalog mapping the distribution of galaxies out to $\simeq 115~$Mpc, the flux-weighted dipole would point in Galactic coordinates in the direction $(\alpha,\delta) = (154^\circ,-9^\circ)$, $\simeq 55^\circ$ away from the signal position. The separation between the two dipole positions is reduced to within $\simeq 2\sigma$ once UHECRs are propagated through models of the Galactic magnetic field.\cite{PierreAuger:2017pzq,PierreAuger:2018zqu} Overall, the signal provides observational evidence that UHECRs, at least above $8~$EeV, are of extragalactic origin.

At the highest energies, magnetic deflections are expected to be less important. Arrival directions can show patterns in a more evident way. Out of all the searches performed, some evidences for anisotropies that mirror to some extent the inhomogeneous distribution of the nearby extragalactic matter have been captured. Above $\simeq~38~$EeV, the flux pattern expected from starburst galaxies (jetted AGN) within 200~Mpc yields the most significant deviation from isotropy at 4.2$\sigma$ (3.3$\sigma$) significance.\cite{PierreAuger:2022axr} The main features of the UHECR arrival directions are shown in Galactic coordinates in Fig.~\ref{fig:largescale} (right panel) by filtering the data at $25^\circ$ above 40~EeV.\cite{PierreAuger:2022axr}

\section{Astrophysical picture}

\begin{wrapfigure}{R}{8 cm}
{\includegraphics[width=0.5\textwidth]{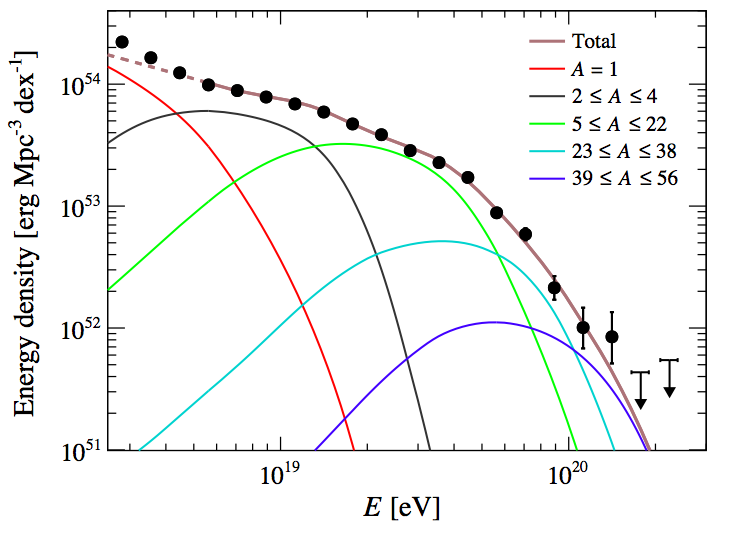}}
\caption{Energy density obtained with the best fit parameters of the benchmark scenario described in the text.}
\label{fig:E2}
\end{wrapfigure}

These results radically changed the perception of UHECRs when the Observatory was designed in the early 1990s. They correspond to a situation in which the total intensity is shaped by a number of different nuclear elements and in which the energy of the nuclei attains a maximum in proportion, at first order, to their charge ($Z$). Namely, the majority of the nuclear elements in the sources are intermediate-mass ones with a maximum energy of $E^{\mathrm{max}}_{Z}=(5\times Z)~$EeV, ranging from He to Si. According to this scenario, the ejection spectrum of the particles that escape from source surroundings is characterized by an extremely hard spectral index that is itself dictated by the quasi-monoelemental increase in average mass with energy observed on Earth.\cite{PierreAuger:2016use} The hardness of the spectral index may be a reflection of how interactions within the sources shaped the ejected spectra of nuclei, which ranged from He to Si or Fe.\cite{Globus:2015xga,Unger:2015laa} According to this picture illustrated in Fig.~\ref{fig:E2}, the maximum energy of acceleration of the heaviest nuclei at the sources and the GZK effect combine to produce the steepening observed above $\simeq 50$~EeV, while the steepening (instep) at $10$~EeV reflects the interplay between the flux contributions of the He and CNO components injected at the source with their distinct cut-off energies, shaped by photo-disintegration during the propagation.\cite{PierreAuger:2020kuy} The substantial flux of protons found below the ankle energy is a direct result of the softer spectral index for protons expected from in-source interactions.\cite{Luce:2022awd} However, it is still unclear where the other nuclear elements that make up the all-particle spectrum up to the ankle feature come from.

\begin{wrapfigure}{L}{8 cm}
{\includegraphics[width=0.5\textwidth]{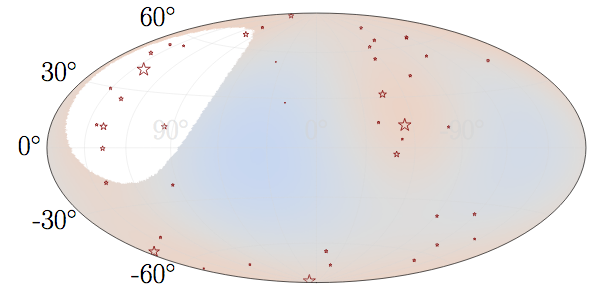}}
\caption{Expected map of UHECR arrival directions above 40 EeV from a model containing a flux contribution of around 20\% of foreground sources from a starburst galaxy catalog (stars). Reddish regions are the hottest ones.}
\label{fig:sbg}
\end{wrapfigure}
The fit of a standard-candle population model to composition and spectral measurements above 1\,EeV can therefore constrain the production rate density of UHECRs. In addition, the arrival directions beyond 40\,EeV can be reproduced by injecting the resulting composition and flux at the sources in the ${\sim}\,$50 brightest star-forming galaxies within 250\,Mpc.\cite{PierreAuger:2023htc} Indeed a model containing a flux contribution from the starburst galaxy catalog of around 20\% at 40 EeV with a magnetic-field blurring angle of around 20$^\circ$ for a rigidity of 10 EV provides a fair simultaneous description of all three observables. The starburst galaxy model turns out to be favored with a significance of 4.5$\sigma$ compared to a  reference model with only homogeneously  distributed sources. As illustrated in Fig.~\ref{fig:sbg}, the Centaurus region provides the dominant contribution to the observed anisotropy signal. Overall, the ring is closing around nearby astrophysical sites to explain the main features of the UHECR flux sky map shown in Fig.~\ref{fig:largescale} (right panel). Yet, the underlying sources traced here by the starburst galaxy catalog remain to be uncovered.

\section{Multi-messenger astronomy and BSM physics from sensitivity to photons and neutrinos} 

\begin{figure}[h]
\centering
\includegraphics[width=0.41\textwidth]{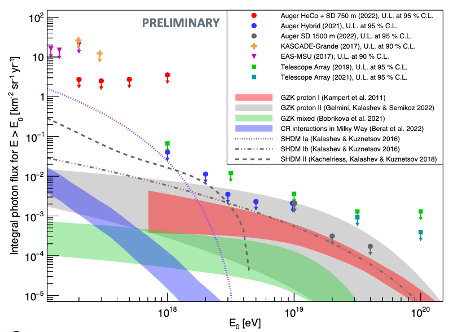}
\includegraphics[width=0.39\textwidth]{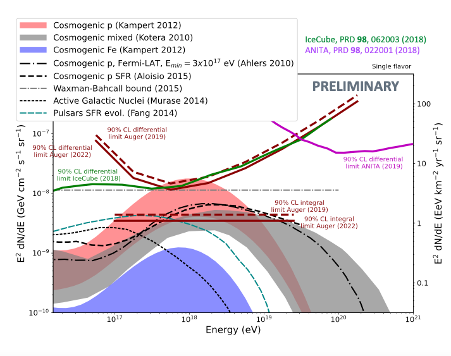}
\caption{Left: 95\% C.L. upper limits to the diffuse flux of UHE photons  (filled circles), shown together with results from other experiments and predictions from several top-down and cosmogenic photon models. Right:  90\% C.L. upper limits (red lines) to the diffuse flux of UHE neutrinos (single flavor) in integrated (horizontal lines) and differential forms, compared with cosmogenic neutrino models, the Waxman-Bahcall bound, and limits from IceCube and ANITA.}
\label{fig:photons_neutrinos}
\end{figure}

Besides the different wavelengths of traditional astronomy, neutrinos, cosmic rays, very high-energy gamma rays and gravitational waves provide complementary information to study the most energetic objects of the Universe. The final boost to the multi-messenger astronomy took place quite recently with the emergence of both neutrino and gravitational-wave astronomy. The discovery of gravitational waves from the merging of a neutron star binary system by LIGO and Virgo triggered a spectacular series of observations in the full electromagnetic spectrum from radio to the very high-energy gamma rays and searches for neutrino fluxes with ANTARES, IceCube, and the Pierre Auger Observatory. The combined effort marks an unprecedented leap forward in astrophysics, revealing many aspects of the gamma-ray burst induced by the merger and its subsequent kilonova. It is now quite clear that there is much potential in the combined analysis of data from different experiments and multi-messenger astronomy has excellent prospects. 

Multi-messenger observations at the Observatory were foreseen from the beginning, in terms of unprecedented sensitivities to ultra-high energy photons and neutrinos (and neutrons). Observations of photons or neutrinos in the so far dark Universe beyond $100~$PeV would be a major breakthrough by itself. Yet, the non-observation of point sources and diffuse fluxes of photons and neutrinos allowed the derivation of upper bounds that constrain models very effectively. The bounds to diffuse photon and neutrino fluxes, shown in Fig.~\ref{fig:photons_neutrinos}, have ruled out the top-down models of UHECR origin motivated by particle physics and also started to exclude,  independently from $X_{\mathrm{max}}$ analyses, the pure proton paradigm. Many TeV gamma-ray sources are observed at energy fluxes of the order of 1~eV~cm$^{-2}$~s$^{-1}$. Such sources would be visible to the Observatory as strong photon and Galactic neutron sources if their energy spectrum would continue with a Fermi-like energy distribution up to about $100~$PeV. Their absence suggests that their maximum source energy does not reach out to the threshold energy of the Observatory and/or that their spectrum is significantly softer than $E^{-2}$.

Point source searches now routinely include also mergers of compact binaries alerted by gravitational wave interferometers. The most spectacular event so far was the neutron star merger GW170817 at a distance of about 40~Mpc.\cite{LIGOScientific:2017zic} Within the predefined $\pm$500~s search window, the  Observatory reached a neutrino flux sensitivity above $100~$PeV that was over an order of magnitude better than of any other neutrino observatory presently operated.~\cite{ANTARES:2017bia} The absence of neutrinos at Auger, IceCube and ANTARES allowed constraining the jet properties of the neutron star merger. Many more events are being expected in the near future. To accommodate for this, mechanisms are set up to automatically react to gravitational wave or other astrophysical alerts and search for neutrinos and photons.

Photons and neutrinos also provide the possibility to constrain models of super-heavy dark matter. This topics is addressed in another proceedings of the conference.

Besides, the fluorescence detector is sensitive to upward-going EASs that would be indicative of BSM physics with possibly new sterile neutrino messengers. To this end, a search for upward-going EASs has been performed. One event has passed the selection criteria, consistent with the expected background ($0.27\pm 0.12$ events).\cite{PierreAuger:2021gci} Upper limits derived from the analysis exclude that the ``anomalous'' ANITA events could signal the emergence of BSM physics.

\section{Puzzles in extensive air showers}

Discrepancies between the muon number $N_\mu$ quantities estimated at the ground level and those predicted by Monte-Carlo generators of EASs have been observed.\cite{PierreAuger:2016nfk} Other discrepancies in $\langle X_{\rm max} \rangle$ have also been uncovered.\cite{PierreAuger:2021xah} 

The interpretation of $X_{\rm max}$ and $N_\mu$ quantities relies on the comparison of their measured values with those predicted by shower simulations, which are resorting to hadronic interaction properties at very high energies and in phase-space regions not well covered by accelerator experiments. Currently, the systematic uncertainties in $X_{\rm max}$ and in $N_\mu$-related quantities predicted by these simulations are dominated by the differences between hadronic interaction models, even after recent updates based on LHC data on parameters governing the interactions. The muon content of showers stems from a multi-step cascade process, mostly driven by interactions of secondary pions and kaons with air. $N_\mu$-related quantities thus depend on properties of pion-air collisions over a wide range of energies, for which a detailed knowledge is lacking. By contrast, $X_{\rm max}$ depends more strongly on the properties of the primary particle interaction with air nuclei. Here, the inelastic cross section and the forward spectra of the secondary hadrons play a key role. In this regard, models maximally benefit from the studies of proton-proton and proton-nucleus collisions at the LHC. An internally consistent hadronic interaction model should reproduce simultaneously the different facets of the showers captured in every constraining observable sensitive to the primary mass. The future insight in the composition-sensitive parameters as a function of the energy from the upgraded Observatory will provide new light on the hadronic processes at energies above those which are achievable with current particle accelerators.

\section{And also...} 

Atmospheric electricity phenomena such as ELVES are being studied since 2013 with the 24 telescopes of the Observatory by exploiting a dedicated trigger and extended readout. The high time resolution allows us to do detailed studies on multiple ELVES. A fraction of multi-ELVES has been uncovered to originate from halos, which are  disc-shaped light transients emitted at 70-80~km altitudes, that follow ELVES after a strong lightning event.\cite{PierreAuger:2021ecs}

Other atmospheric electricity phenomena studied at the Observatory are downward Terrestrial Gamma-ray Flashes, which are millisecond bursts of gamma-rays originating from within the Earth’s atmosphere during thunderstorm. They happen in coincidence with low thunderclouds and lightning, and their large deposited energy at the ground is compatible with the source a few kilometers above the ground. A few handful of events have been recorded with the particle detectors. These events can give important clues about the production models.\cite{PierreAuger:2021int}

Finally, the Pierre Auger Collaboration has embraced the concept of open access to their data. A new portal was opened in February 2021, at the end of the first phase of operation. It includes 10\% of the data -- raw and processed -- from the different instruments of the Observatory, a visualisation tool, documentation to make the data user-friendly, and analyses codes that can be readily used and modified. The portal has become dual, with one part dedicated to scientists and the other to educational users. Furthermore, a catalog containing details of the 100 highest-energy cosmic rays has been included.\cite{PierreAuger:2022qcg}

\section*{References}

\bibliography{biblio}

\begin{thebibliography}{10}

\bibitem{PierreAuger:2015eyc}
The Pierre~Auger Collaboration.
\newblock {The Pierre Auger Cosmic Ray Observatory}.
\newblock {\em Nucl. Instrum. Meth. A}, 798:172--213, 2015.

\bibitem{PierreAuger:2014sui}
The Pierre~Auger Collaboration.
\newblock {Depth of Maximum of Air-Shower Profiles at the Pierre Auger
  Observatory: Measurements at Energies above $10^{17.8}$ eV}.
\newblock {\em Phys. Rev. D}, 90(12):122005, 2014.

\bibitem{PierreAuger:2008rol}
The Pierre~Auger Collaboration.
\newblock {Observation of the suppression of the flux of cosmic rays above
  $4\times 10^{19}$eV}.
\newblock {\em Phys. Rev. Lett.}, 101:061101, 2008.

\bibitem{PierreAuger:2021hun}
The Pierre~Auger Collaboration.
\newblock {The energy spectrum of cosmic rays beyond the turn-down around
  $10^{17}$ eV as measured with the surface detector of the Pierre Auger
  Observatory}.
\newblock {\em Eur. Phys. J. C}, 81(11):966, 2021.

\bibitem{PierreAuger:2020qqz}
The Pierre~Auger Collaboration.
\newblock {Measurement of the cosmic-ray energy spectrum above $2.5{\times}
  10^{18}$ eV using the Pierre Auger Observatory}.
\newblock {\em Phys. Rev. D}, 102(6):062005, 2020.

\bibitem{PierreAuger:2019phh}
{\em {The Pierre Auger Observatory: Contributions to the 36th International
  Cosmic Ray Conference (ICRC 2019)}: {Madison, Wisconsin, USA, July 24- August
  1, 2019}}, 9 2019.

\bibitem{PierreAuger:2014gko}
The Pierre~Auger Collaboration.
\newblock {Depth of maximum of air-shower profiles at the Pierre Auger
  Observatory. II. Composition implications}.
\newblock {\em Phys. Rev. D}, 90(12):122006, 2014.

\bibitem{PierreAuger:2020fbi}
The Pierre~Auger Collaboration.
\newblock {Cosmic-ray anisotropies in right ascension measured by the Pierre
  Auger Observatory}.
\newblock {\em Astrophys. J.}, 891:142, 2020.

\bibitem{Edge:1978rr}
D.~M. Edge, A.~M.~T. Pollock, R.~J.~O. Reid, A.~A Watson, and J.~G. Wilson.
\newblock {A Study of the Arrival Direction Distribution of High-Energy
  Particles as Observed from the Northern Hemisphere}.
\newblock {\em J. Phys. G}, 4:133--157, 1978.

\bibitem{Deligny:2018blo}
Olivier Deligny.
\newblock {Measurements and implications of cosmic ray anisotropies from TeV to
  trans-EeV energies}.
\newblock {\em Astropart. Phys.}, 104:13--41, 2019.

\bibitem{PierreAuger:2017pzq}
The Pierre~Auger Collaboration.
\newblock {Observation of a Large-scale Anisotropy in the Arrival Directions of
  Cosmic Rays above $8 \times 10^{18}$ eV}.
\newblock {\em Science}, 357(6537):1266--1270, 2017.

\bibitem{PierreAuger:2018zqu}
The Pierre~Auger Collaboration.
\newblock {Large-scale cosmic-ray anisotropies above 4 EeV measured by the
  Pierre Auger Observatory}.
\newblock {\em Astrophys. J.}, 868(1):4, 2018.

\bibitem{PierreAuger:2022axr}
The Pierre~Auger Collaboration.
\newblock {Arrival Directions of Cosmic Rays above 32 EeV from Phase One of the
  Pierre Auger Observatory}.
\newblock {\em Astrophys. J.}, 935(2):170, 2022.

\bibitem{PierreAuger:2016use}
The Pierre~Auger Collaboration.
\newblock {Combined fit of spectrum and composition data as measured by the
  Pierre Auger Observatory}.
\newblock {\em JCAP}, 04:038, 2017.
\newblock [Erratum: JCAP 03, E02 (2018)].

\bibitem{Globus:2015xga}
Noemie Globus, Denis Allard, and Etienne Parizot.
\newblock {A complete model of the cosmic ray spectrum and composition across
  the Galactic to extragalactic transition}.
\newblock {\em Phys. Rev. D}, 92(2):021302, 2015.

\bibitem{Unger:2015laa}
Michael Unger, Glennys~R. Farrar, and Luis~A. Anchordoqui.
\newblock {Origin of the ankle in the ultrahigh energy cosmic ray spectrum, and
  of the extragalactic protons below it}.
\newblock {\em Phys. Rev. D}, 92(12):123001, 2015.

\bibitem{PierreAuger:2020kuy}
The Pierre~Auger Collaboration.
\newblock {Features of the Energy Spectrum of Cosmic Rays above
  2.5\texttimes{}10$^{18}$ eV Using the Pierre Auger Observatory}.
\newblock {\em Phys. Rev. Lett.}, 125(12):121106, 2020.

\bibitem{Luce:2022awd}
Quentin Luce, Sullivan Marafico, Jonathan Biteau, Antonio Condorelli, and
  Olivier Deligny.
\newblock {Observational Constraints on Cosmic-Ray Escape from Ultrahigh-energy
  Accelerators}.
\newblock {\em Astrophys. J.}, 936(1):62, 2022.

\bibitem{PierreAuger:2023htc}
The Pierre~Auger Collaboration.
\newblock {Constraining models for the origin of ultra-high-energy cosmic rays
  with a novel combined analysis of arrival directions, spectrum, and
  composition data measured at the Pierre Auger Observatory}.
\newblock {\em JCAP}, submitted, 2023.

\bibitem{LIGOScientific:2017zic}
B.~P. Abbott et~al.
\newblock {Gravitational Waves and Gamma-rays from a Binary Neutron Star
  Merger: GW170817 and GRB 170817A}.
\newblock {\em Astrophys. J. Lett.}, 848(2):L13, 2017.

\bibitem{ANTARES:2017bia}
A.~Albert et~al.
\newblock {Search for High-energy Neutrinos from Binary Neutron Star Merger
  GW170817 with ANTARES, IceCube, and the Pierre Auger Observatory}.
\newblock {\em Astrophys. J. Lett.}, 850(2):L35, 2017.

\bibitem{PierreAuger:2021gci}
The Pierre~Auger Collaboration.
\newblock {Search for upward-going showers with the Fluorescence Detector of
  the Pierre Auger Observatory}.
\newblock {\em PoS}, ICRC2021:1140, 2021.

\bibitem{PierreAuger:2016nfk}
The Pierre~Auger Collaboration.
\newblock {Testing Hadronic Interactions at Ultrahigh Energies with Air Showers
  Measured by the Pierre Auger Observatory}.
\newblock {\em Phys. Rev. Lett.}, 117(19):192001, 2016.

\bibitem{PierreAuger:2021xah}
Jakub Vicha et~al.
\newblock {Adjustments to Model Predictions of Depth of Shower Maximum and
  Signals at Ground Level using Hybrid Events of the Pierre Auger Observatory}.
\newblock {\em PoS}, ICRC2021:310, 2021.

\bibitem{PierreAuger:2021ecs}
Adriana V\'asquez~Ram\'\i{}rez et~al.
\newblock {Study on multi-ELVES in the Pierre Auger Observatory}.
\newblock {\em PoS}, ICRC2021:327, 2021.

\bibitem{PierreAuger:2021int}
Roberta Colalillo et~al.
\newblock {Downward Terrestrial Gamma-ray Flashes at the Pierre Auger
  Observatory?}
\newblock {\em PoS}, ICRC2021:395, 2021.

\bibitem{PierreAuger:2022qcg}
The Pierre~Auger Collaboration.
\newblock {A Catalog of the Highest-energy Cosmic Rays Recorded during Phase I
  of Operation of the Pierre Auger Observatory}.
\newblock {\em Astrophys. J. Suppl.}, 264(2):50, 2023.

\end{thebibliography}

\end{document}